\documentclass[a4paper,11pt]{article}

\usepackage{contribution}

\newcommand{\weblink}[2][]{%
    \ifthenelse{\equal{#1}{}}%
    {\textnormal{\url{#2}}}%
    {\textnormal{\href{#2}{#1}}}%
}

\def\beq{\begin{equation}}
\def\eeq#1{\label{#1}\end{equation}}
\def\eeqn{\end{equation}}

\def\beqa{\begin{eqnarray}}
\def\eeqa#1{\label{#1}\end{eqnarray}}
\def\eeqan{\end{eqnarray}}

\let\bar=\overbar

\def\Dslash{\not{\hbox{\kern-4pt $D$}}}
\def\dslash{\not{\hbox{\kern-2pt $\del$}}}

\def\msb{{\bar{\ssstyle M \kern -1pt S}}}

\newcommand{\contribution}[7][]{%
  \clearpage
  \thispagestyle{plain}
  \ifthenelse{\equal{#1}{}}
  {\hypersetup{pdftitle={#2}}}
  {\hypersetup{pdftitle={#1}}}
  \hypersetup{pdfauthor={{#3} {#4}}}
  {\centering\normalfont\LARGE\bfseries\sffamily #2 \par\nobreak}
  \lhead{}
  \chead{%
    \textit{\footnotesize XIV International Conference on Hadron Spectroscopy
      (\weblink[\textit{hadron2011}]{http://www.hadron2011.de}), 13-17 June 2011, Munich, Germany}%
  }
  \rhead{}
  \bigskip
  \begin{center}
    {#3} {#4}\ifthenelse{\equal{#6}{}}{}{\footnote{\weblink[#6]{mailto:#6}}}
    \ifthenelse{\equal{#7}{}}{}{#7} \\
    \textit{#5}
  \end{center}
  \bigskip
}

\renewcommand{\abstract}[1]{%
  \begin{center}
    \begin{minipage}{0.85\textwidth}
      \begin{footnotesize}
        #1
      \end{footnotesize}
    \end{minipage}
  \end{center}
  \bigskip
}

\begin{document}

%
%
%
%
%
{  

\makeatletter
\@ifundefined{c@affiliation}%
{\newcounter{affiliation}}{}%
\makeatother
\newcommand{\affiliation}[2][]{\setcounter{affiliation}{#2}%
  \ensuremath{{^{\alph{affiliation}}}\text{#1}}}
%

%

\contribution[$\eta - \eta^{\prime}$ Mixing -- From Electromagnetic Transitions to Weak Decays of
Charm and Beauty Hadrons]
{$\eta - \eta^{\prime}$ Mixing -- From Electromagnetic Transitions \\ to Weak Decays of
Charm and Beauty Hadrons}
{Camilla}{Di Donato}  
{\affiliation[I.N.F.N. Sezione di Napoli,
Complesso M.S.A. --- via Cintia, 80126 Napoli, Italy ]{1} \\
 \affiliation[Dipartimento di Scienze Fisiche, Universit\'a di Napoli Federico II,
Complesso M.S.A.]{2} \\
 \affiliation[Department of Physics, University of Notre Dame du Lac,
Notre Dame, IN 46556, USA]{3}}
{camilla.didonato@na.infn.it, giulia.ricciardi@na.infn.it, ibigi@nd.edu }
{\!\!$^,\affiliation{1}$, Giulia Ricciardi\affiliation{1}$^,$\affiliation{2}, Ikaros Bigi\affiliation{3}}
%

\abstract{%
It has been realized for a long time that knowing the $\eta$ and $\eta^{\prime}$ wave functions in terms of quark and gluon components probes our understanding of non-perturbative QCD dynamics. Great effort has been given to this challenge -- yet no clear picture has emerged even with the most recent KLOE data. We point out which measurements would be most helpful in arriving at a more definite conclusion. A better knowledge of these wave functions will significantly help to disentangle the weight of different decay subprocesses in semi-leptonic decays of $D^+$, $D_s^+$ and $B^+$ mesons. The resulting insights will be instrumental in treating even non-leptonic B transitions involving $\eta$ and $\eta^{\prime}$ and their CP asymmetries; thus they can sharpen the case for or against New Physics intervening there.}
%

\section{Introduction}
The question of the $\eta - \eta^{\prime}$ mixing go back to the beginning of the quark model era and with the advent of QCD it became even more involved, since QCD brought with it more dynamical degrees of freedom, the gluons, which can form a second class of $SU(3)$ singlet. Showing there is  a pure gluonic component would establish for the first time that gluons play an independent role also in hadronic spectroscopy; we know that the presence of gluons as independent degree of freedom has been demonstrated as progenitors of jets in hard collisions.
Knowing $\eta - \eta^{\prime}$ mixing is essential also to disentangle SM hadronic uncertanties versus New Physics. Great effort has been given to this challenge, but yet not clear picture has emerged\footnote{The complete bibliography can be found in reference \cite{DiDonato:2011kr}} \cite{DiDonato:2011kr}.\\ 
In this paper we retrace the essential element to treat the $\eta-\eta^\prime$ mixing, starting from the electromagnetic processes up to weak decays of charm and beauty hadrons to search for a thread for more definite conclusions.

\subsection{The Pseudoscalar Mixing Angle}
The mixing of the $\eta$ and $\eta^{\prime}$ mesons can be  described in two different bases: the $SU(3)_{fl}$ singlet and octet components, $|\eta_0\rangle$ and $|\eta_8\rangle$, respectively; the quark-flavor basis with
$|\eta_{q}\rangle$ and $|\eta_{s}\rangle$:
\begin{eqnarray*}
\left (
\begin{array}{c}
|\eta \rangle \\
|\eta^{\prime} \rangle \\
\end{array}
\right ) =
\left (
\begin{array}{cc}
\cos \theta_P & -\sin \theta_P \\
\sin \theta_P & \cos \theta_P \\
\end{array}
\right )
\left (
\begin{array}{c}
|\eta_8 \rangle \\
|\eta_0 \rangle \\
\end{array}
\right )  \; 
\rightarrow
\left (
\begin{array}{cc}
\cos \phi_P & -\sin \phi_P \\
\sin \phi_P & \cos \phi_P \\
\end{array}
\right )
\left (s
\begin{array}{c}
|\eta_q \rangle \\
|\eta_{s} \rangle \\
\end{array}
\right ) 
\end{eqnarray*}
Just for orientation: the quadratic [linear] Gell-Mann Okubo (GMO) mass formula points to $\theta_P\simeq -10^{\circ}$,
$\phi_P\simeq 44.7^{\circ}$
[$\theta_P\simeq -23^{\circ}$, $\phi_P\simeq 31.7^{\circ}$].

In the nineties several papers \cite{Feldmann:1999uf} have been shown the mixing cannot be adequately described by a single angle and, due to $SU(3)_{fl}$ breaking, the mixing of decay constants does not necessarily follow the same pattern as state mixing. We have two mixing angle: $\theta_8$ and $\theta_0$ in the $SU(3)_{fl}$ and $\phi_q$ and $\phi_s$ in the flavour basis.
The estimated difference $\theta_8-\theta_0$ can be large as 
$-12^{\circ}/-19^{\circ}$.
The quark-flavour basis play a key rule, indeed 
the mixing is large, order $\simeq 40^{\circ}$, but the difference between the two mixing angles is determined by OZI-rule violating contribution: the angles $\phi_q-\phi_s$ nearly coincide.

In QCD we can have a $SU(3)_{fl}$ singlet from quark-antiquark combinations and also from pure gluon configurations with the simplest one being a gluon-gluon (gg) combination. We take into account the possibility that the $\eta$ and $\eta^{\prime}$ wave functions could contain such configuration, gluonium. We will ignore mixing with heavier pseudoscalar mesons, because the scale for gluonic excitations is presumably lower then the $J/\Psi$ mass.
One would also expect the heavier $\eta^{\prime}$ to contain a higher dose of gluonic component then the $\eta$, which is also mainly a $SU(3)_{fl}$ octet.
Present evaluations come from phenomenological analyses.

\section{Electromagnetic and Strong Transitions}
Several electromagnetic and strong transitions can provide information on the mixing angles and the gluonic content: radiative vector and pseudoscalar meson decays
$ \psi^\prime, \, \psi , \,  \phi  \to \gamma \eta^{\prime} \;  vs. \;  \gamma \eta  \;$;  $  \rho, \, \omega  \to \gamma \eta$ and $ \eta^{\prime} \to \gamma    \omega,  \, \gamma \rho$; 
decays into two photons or production in $\gamma\gamma$ collisions, $ \eta^{\prime} \to 2\gamma$ vs. $ \eta \to 2\gamma $, $ \; \gamma \gamma \to  \eta \;$ vs. $ \; \gamma \gamma \to  \eta^{\prime} $; decays of $\psi$ into PV final states with the vector meson acting as a `flavour filter',
$\psi \to \rho /\omega /\phi + \eta \;$ vs. $\; \eta^{\prime}$.
The modern analyses \cite{Ambrosino:2006gk,Escribano:2007cd,Thomas:2007uy} get consistent results, but their message, concerning the gluonium content of $\eta^\prime$, is ambivalent. The conclusion related to the discrepancy among those analyses is that the $\eta^\prime \to \gamma\gamma$ seems to play a key role in the mixing parameters  determination and the inverse process, $\gamma\gamma \to \eta^\prime$, provides an important check of whether a resonance is a conventional $q \bar q$.  

\section{Weak Decays of Charm and Beauty Hadrons}
Knowing reliably the $\eta$ and $\eta^\prime$ wave functions is an important input for our understanding of several weak decays of beauty and charm hadrons. Most crucially we need it for predicting of CP asymmetries involving $\eta$ and $\eta^\prime$ in the final states and to understand whether a deviation from SM predictions can be seen as a signal of New Physics.
The approach in the analysis is the phenomenological one, but lattice QCD simulation have just entered the adult period and first calculation gives $\phi_P$ order $40^\circ$.
Since one expects semi-leptonic transitions to be driven by SM dynamics only, their detailed studies teach us lessons on how non-perturbative hadronization transforms quark level transitions.  
The transitions $D_s^+ \to \eta^{(\prime )} l^+ \nu$, $D^+\to \eta^{(\prime )} l^+ \nu$ and $B^+ \to \eta^{(\prime )} l^+ \nu$ proceed on greatly different time scales, since they are driven by weak interactions on the Cabibbo allowed, Cabibbo suppressed and KM suppressed levels, respectively. Yet they can provide us with highly complementary information in the sense that
they produce the $\eta^{(\prime )}$ via their $s \bar s$, $d \bar d$ and $u \bar u$ components, respectively. The analyses of those processes again indicate a mixing angle $\phi_P$ order $40^\circ$. Considering also the fact that $\eta^{(\prime )}$ could be excited via a $gg$ component 
the available data indicate, with a $\phi_P=37.7^\circ$, an
$\eta^\prime$ gluonium content at level of $12\%$ \cite{DiDonato:2011kr}.
The theoretical situation is more complex while spectator diagrams generate leading contributions. The so-called weak annihilation (WA) process contributes moderately to inclusive decay rates, but it could affect exclusive channel considerably.
The strength of the effect depends on the size of the gluonic component in the $\eta^\prime$ wave function and on how much gg radiation one can expect in semileptonic $D_s^+$, $D^+$ and $B^+$ decays. Since the main effect might come from the interference with the spectator amplitude, it can a priori enhance or reduce those rates. Recent analyses \cite{Gambino:2010jz}, based on exclusive semileptonic $D$ decays, which considers both the widths and the lepton energy moments, shows no clear evidence of WA effects.
In $B^\pm$ semileptonic decays with $\eta-\eta^\prime$ in the final states the situation is analogous to that for the $D^\pm$, except that their rates are suppressed by $|V_{us}/V_{cb}|^2$ rather than $|V_{cd}/V_{cs}|^2$ and that the range in $q^2$ is much large. CLEO \cite{Adam:2007pv} gives first evidence of $B^+ \to \eta l^+ \nu$, and later on also BaBar \cite{:2008gka} has mesured the branching ratio, but the experimental situation is not yet satisfying: anyway the ratio
${\cal B}(B^+ \rightarrow  \eta^\prime l^+ \nu)/{\cal B}(B^+ \rightarrow  \eta l^+ \nu) = 0.67 \pm 0.24_{stat} \pm 0.11_{syst}$ \cite{delAmoSanchez:2010zd} seems to allow  a large gluonic singlet contribution to the $\eta^\prime$ form factor.
The corresponding ratio involving the $B_s$ mesons,
${\cal B}(B_s \rightarrow  \eta^\prime l^+ l^-)/{\cal B}(B_s \rightarrow  \eta l^+ l^-) $, is also potentially informative on the $\eta^{(\prime)}$ gluonic content, although experimentally much more challenging; they are within the reach of facilities
like the Super-Flavour factories.
\section{CP Violation}
Within the SM many charmless non-leptonic $B$ decays receive significant or even leading contributions
from loop processes, which represent
quantum corrections. Thus they provide fertile hunting grounds for New Physics, in particular in their CP asymmetries. 
Modes like $B \to \eta^{(\prime)}K$ and $B_s \to \eta^{(\prime)}\phi$ seem particularly well-suited in this
respect.
The $B \to \eta^{(\prime)}K$ decays may proceed through tree diagrams $\bar b \to \bar u u \bar s$, but such contributions are color and CKM suppressed; then we have penguins $b \to s$.
The same basic penguin mechanism is expected to drive both $B \to \eta^{(\prime)}K$ and $B \to \pi K$, but the rate of the former is measured to be much larger. A possible distinctive contribution are flavour singlet amplitudes, that are not allowed if the final state contains only flavour non singlet states such pions or kaons. In flavour singlet penguins two gluons couple to the $\eta^\prime$ violating the OZI rule and the amplitude can get contributions from the pure gluonic component of the $\eta^\prime$.
%


\section{Conclusions}

In conclusion, after many and hard efforts to understand the $\eta-\eta^\prime$ wave functions it is still intriguing to improve our understanding of the non perturbative effects in QCD, it will help to identify new physics in CP asymmetries in $B$ and $D$ decays.

%

}  


\end{document}